\documentclass[showpacs,superscriptaddress,preprint,prl]{revtex4}
\usepackage{amsmath,graphicx}
\usepackage{color,ulem}

\definecolor{darkred}{rgb}{0.90,0,0}
\definecolor{darkgreen}{rgb}{0,0.60,.2}
\definecolor{darkblue}{rgb}{0,0,1}
\definecolor{grey}{cmyk}{0,0,0,0.25}
\definecolor{orange}{cmyk}{0,0.6,0.8,0}

\newcommand{\be}{\begin{equation}}
\newcommand{\ee}{\end{equation}}

\begin{document}

\title{Thin film barristor: a gate tunable vertical graphene-pentacene device}
\author{C. Ojeda-Aristizabal}
\affiliation{Center for Nanophysics and Advanced Materials, University of Maryland, College Park, MD 20742-4111, USA}
\author{W. Bao}
\affiliation{Center for Nanophysics and Advanced Materials,
University of Maryland, College Park, MD 20742-4111, USA}
\author{M. S. Fuhrer }
\affiliation{Center for Nanophysics and Advanced Materials, University of Maryland, College Park, MD 20742-4111, USA}
\affiliation{School of Physics, Monash University, Victoria 3800, Australia}

\begin{abstract}
We fabricate a vertical thin-film barristor device consisting of highly doped silicon (gate), 300 nm SiO$_2$ (gate dielectric), monolayer graphene, pentacene, and a gold top electrode. We show that the current across the device is modulated by the Fermi energy level of graphene, tuned with an external gate voltage. We interpret the device current within the thermionic emission theory, showing a modulation of the energy barrier between graphene and pentacene as large as 300meV.
\end{abstract}
\maketitle

Graphene, a one atom thick crystal made of carbon, shows exciting possibilities as a tunable electrode for semiconductors \cite{Lee}, \cite{Barristor}, \cite{Hyoung}, \cite{Cho}, \cite{Wang}, \cite{Berke}. Graphene's electrochemical potential can be tuned over a span of electron volts \cite{Yu}, and it is expected to have no interface states. Here we explore graphene as a tunable electrode contacting pentacene, a van der Waals molecular p-type semiconductor which should also have no interface states. Pentacene has interest in organic electronics due to its low cost fabrication, low temperature processing and mechanical flexibility \cite{Kitamura}, \cite{BookLi}. Here we demonstrate a vertical Si-Si$O_2$-graphene-pentacene-gold barristor device. Gate voltage applied to the silicon modulates the Fermi energy of graphene and controls the current through the vertical graphene-pentacene-gold structure. We observe that the activation barrier for thermionic emission from graphene to pentacene can be tuned by gate voltage up to 300 meV.

Graphene was exfoliated on a substrate of 300 nm SiO$_2$ over highly doped Si and electrical contact to a portion of the graphene was established via deposition of chromium/gold  electrodes through a silicon physical mask \cite{Wenzhong}. The sample was annealed in an Ar atmosphere (~1700mL/min) at 250$^{\circ}$C for 1 hour. Before depositing the pentacene, a negative resist of $~$150nm thick (hydrogen silsesquioxane,HSQ) was applied to the sample by spin coating and baked at 80$^{\circ}$C. A small window of 9x9 $\mu$m$^2$ was opened over the graphene using electron beam lithography followed by developing in MF-26A (2$\%$ tetramethylammonium hydroxide). Pentacene was subsequently deposited; first at a rate of 0.3 $\mathring{A}$/s up to a thickness of 80 nm and then at 2.4 $\mathring{A}/s$ to complete $\sim$780 nm. A thin (50 nm) layer of gold of was deposited at a rate of 3 $\mathring{A}/s$ to establish a top electrical contact to the pentacene. The result is a vertical silicon/SiO$_2$/graphene/pentacene/gold device with three terminals: metal electrodes make independent contacts to graphene and pentacene, and the silicon acts as a third gate terminal. The negative HSQ resist confines the device to a small region of the graphene, and the rest of the graphene electrically shields the pentacene from the gate field; the graphene extends outward from the device area for several microns, much greater than the device thickness $\sim$780 nm. The absence of a direct gate field effect on the pentacene was also verified in devices using thick graphite bottom electrodes; see below. In order to verify that the gate field effect is due to Fermi energy change in the graphene, we have fabricated a control device with a $\sim2\mu$m thick graphite bottom electrode. We expect that, due to the larger areal density of electronic states in the thick graphite that the gate voltage has negligible effect on the Fermi level in the control device.

Atomic force microscopy of the pentacene before top electrode deposition is consistent with previous reports which observed a different structure for pentacene on graphite, interpreted as  pentacene molecules oriented parallel to the graphite layers \cite{Hoshino}, \cite{Hyoung}. This is explained by the alignment of the $\pi$ orbital of graphene and pentacene which gives rise to a better contact resistance compared to other metals contacting pentacene\cite{Sangchul}, \cite{SangchulAPL}.

\begin{figure}[H]
\begin{center}
\begin{tabular}{c c}
\includegraphics[width=8cm]{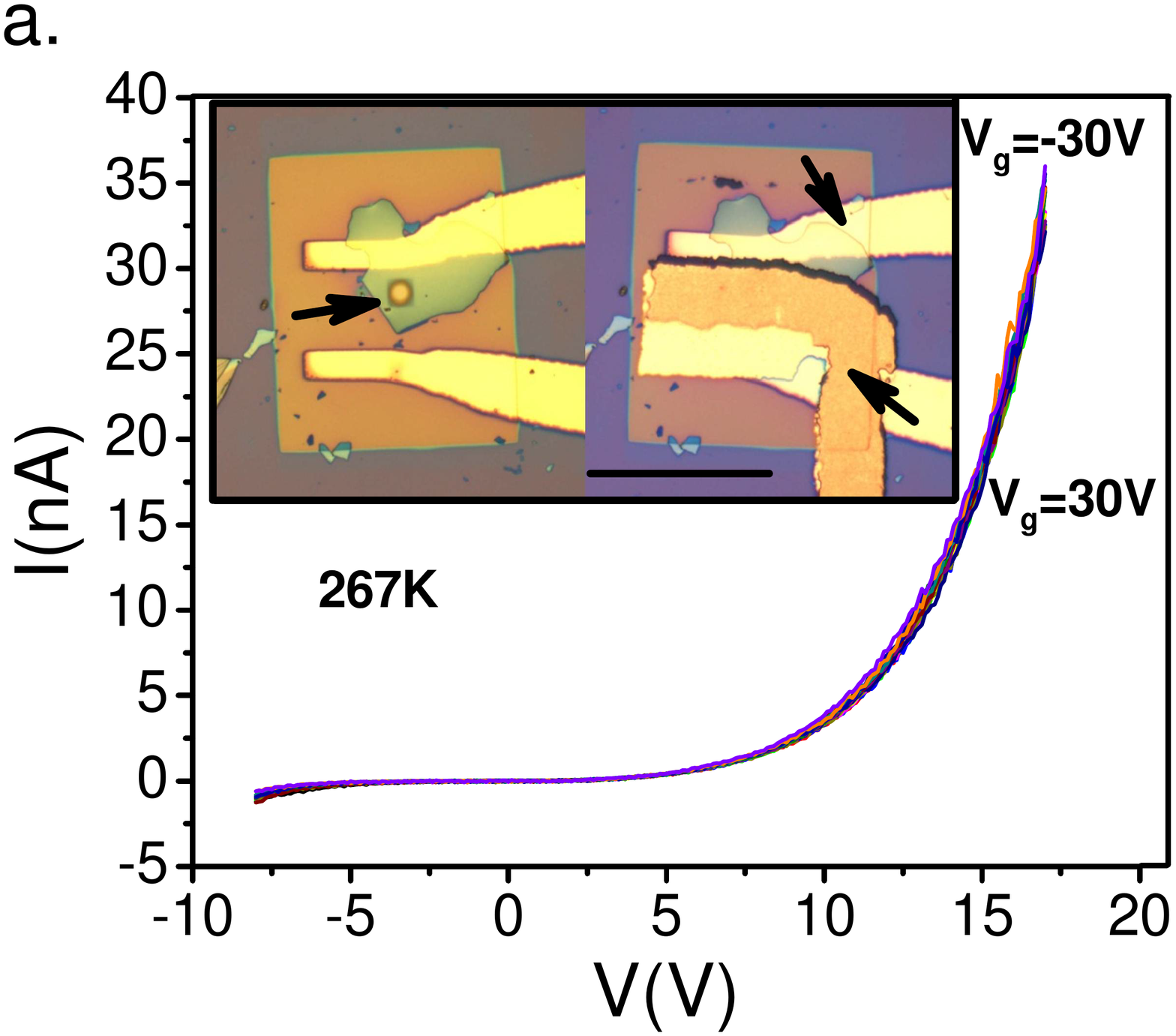}
\includegraphics[width=8cm]{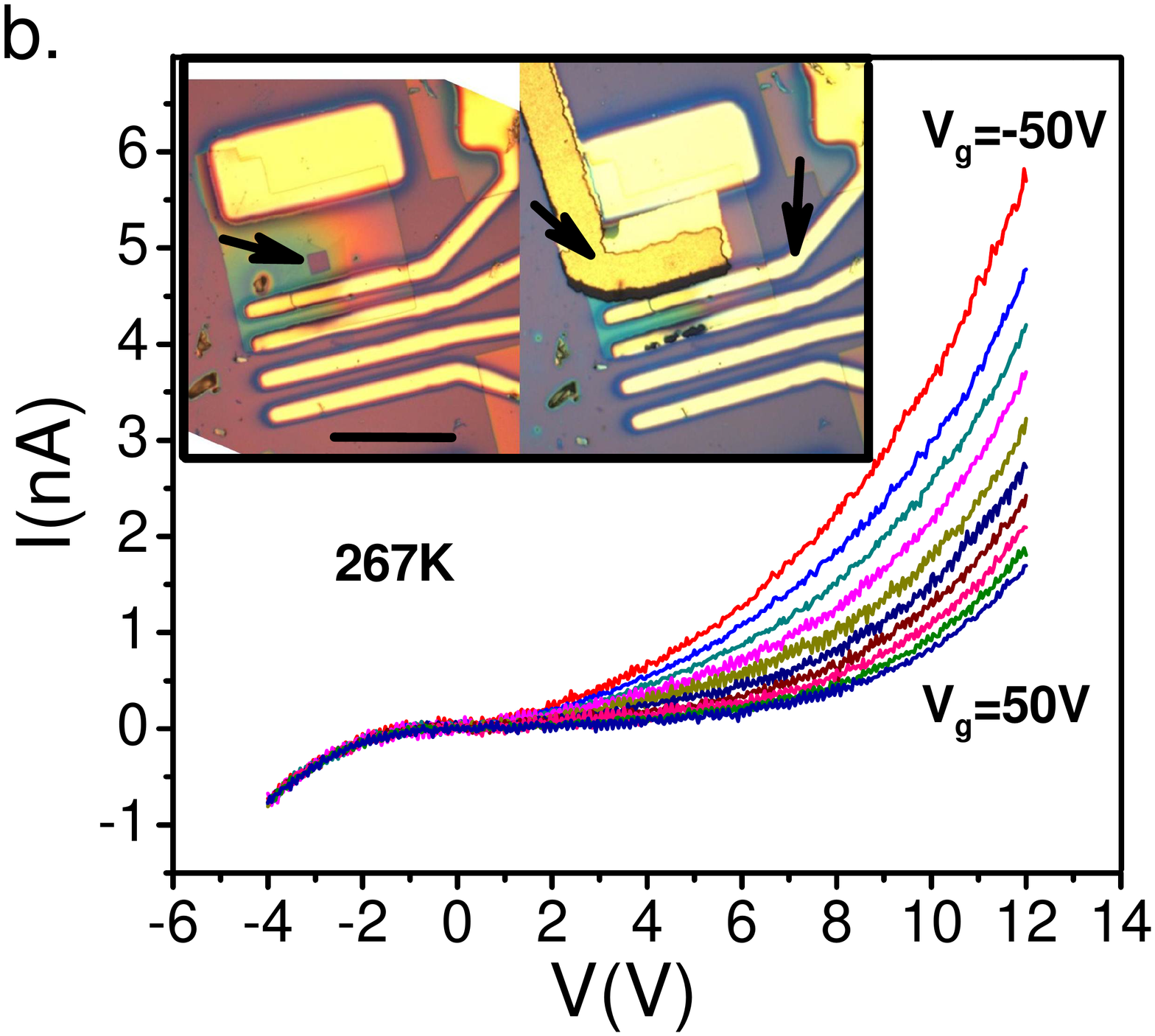}
\end{tabular}
\caption{Current I through the graphite (a) and graphene (b) - pentacene vertical device as a function of the bias voltage V at different gate voltages V$_g$. Insets show the respective optical images of the devices before (left) and after (right) deposition of top electrode. The bar corresponds to 50 $\mu$m. The arrows indicate the window opened on a 150 nm thick HSQ layer (left) and the two electrodes used for transport (right).}
\label{Fig1}
\end{center}
\end{figure}

Fig. 1 shows the current-voltage $I(V)$ characteristics of the devices at different gate voltages $V_g$. In both the graphite and the graphene device, transport was measured between the gold electrode contacting graphite/graphene and the pentacene electrode. The graphite control device, shown in figure \ref{Fig1} a) presents $I(V)$ characteristics that depended only weakly on the gate voltage. However in the graphene device (figure \ref{Fig1} b) the current is strongly modulated by gate voltage (by a factor of 4 over a gate voltage range of 100 V).

To further explore the effect of the gate voltage on the graphene device, measurements were repeated at different temperatures $T$. Figure 2 shows the temperature dependence plotted as ln$(I/T^2)$ as a function of $1/T$ at various gate voltages. Straight lines indicate  thermally activated behavior of the current within the Richardson-Dushman thermionic emission theory \cite{Sze},
\be\label{Thermionic}
J=-A^*T^2exp\big(-\frac{q\phi_B}{k_BT}\big).
\ee
where $\phi_B$ is the energy barrier height and $A^*$ the Richardson constant.

Two regimes of temperature dependence are evident in Fig. 2. At high temperature ($1/T<0.00347$ or $T>288$ K) the slope is relatively independent of gate voltage. However at lower temperature ($1/T>0.00347$ or $T<288$ K) the slope varies strongly with gate voltage, indicating a change in the activation energy for the process controlling transport.

\begin{figure}[H]
\begin{center}
\begin{tabular}{c c}
\includegraphics[width=8cm]{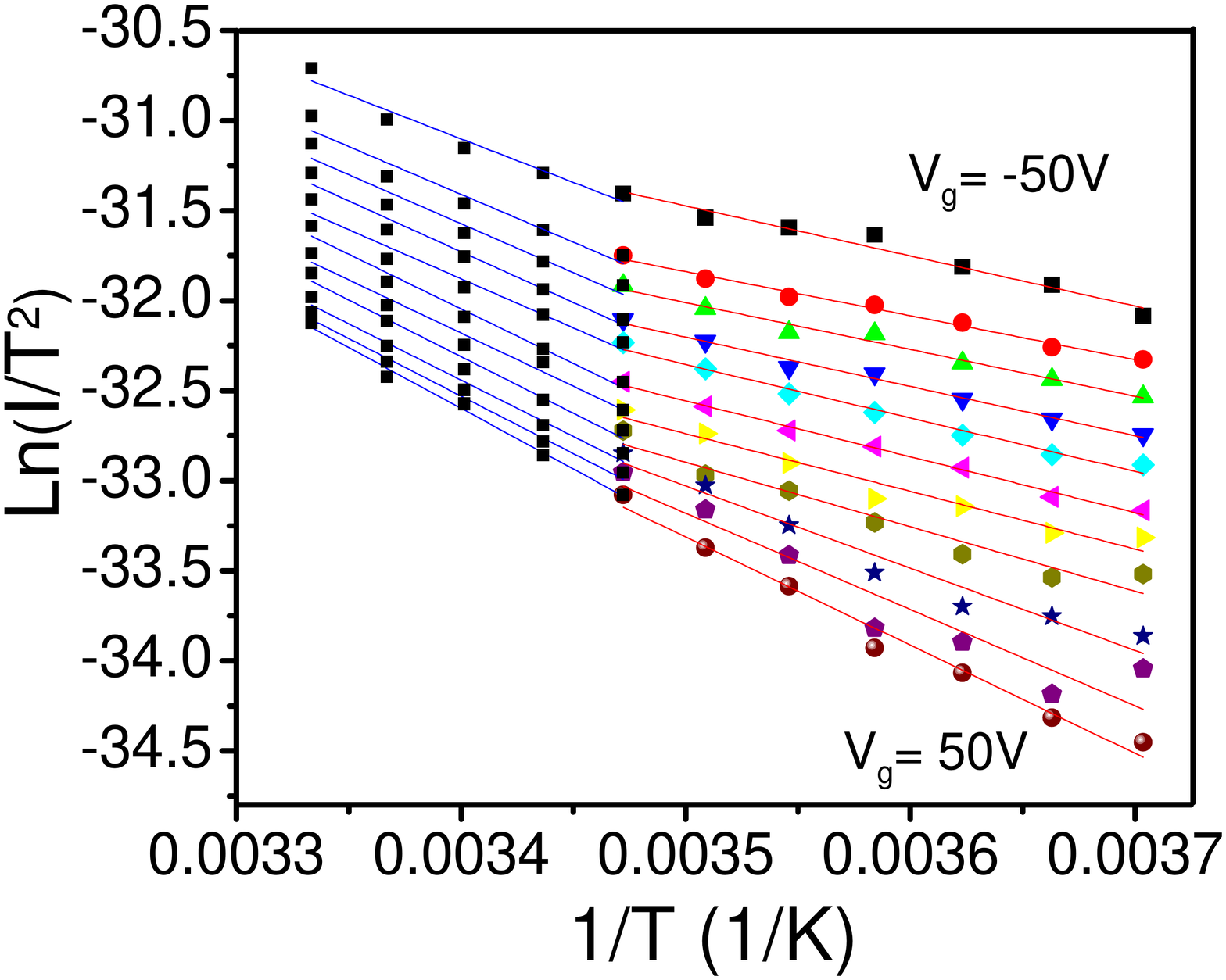}
\end{tabular}
\caption{Semilog plot of I/T$^2$ as a function of 1/T for the graphene-pentacene current at different gate voltages. Current was measured at V$_{DC}$=4V. Straight lines are fits to the Richardson-Dushman thermionic emission theory (equation \ref{Thermionic}). Fits are done in two different temperature ranges: 288K-300K and 270K-288K.}
\label{Fig2}
\end{center}
\end{figure}

\begin{figure}[H]
\begin{center}
\begin{tabular}{c c}
\includegraphics[width=8cm]{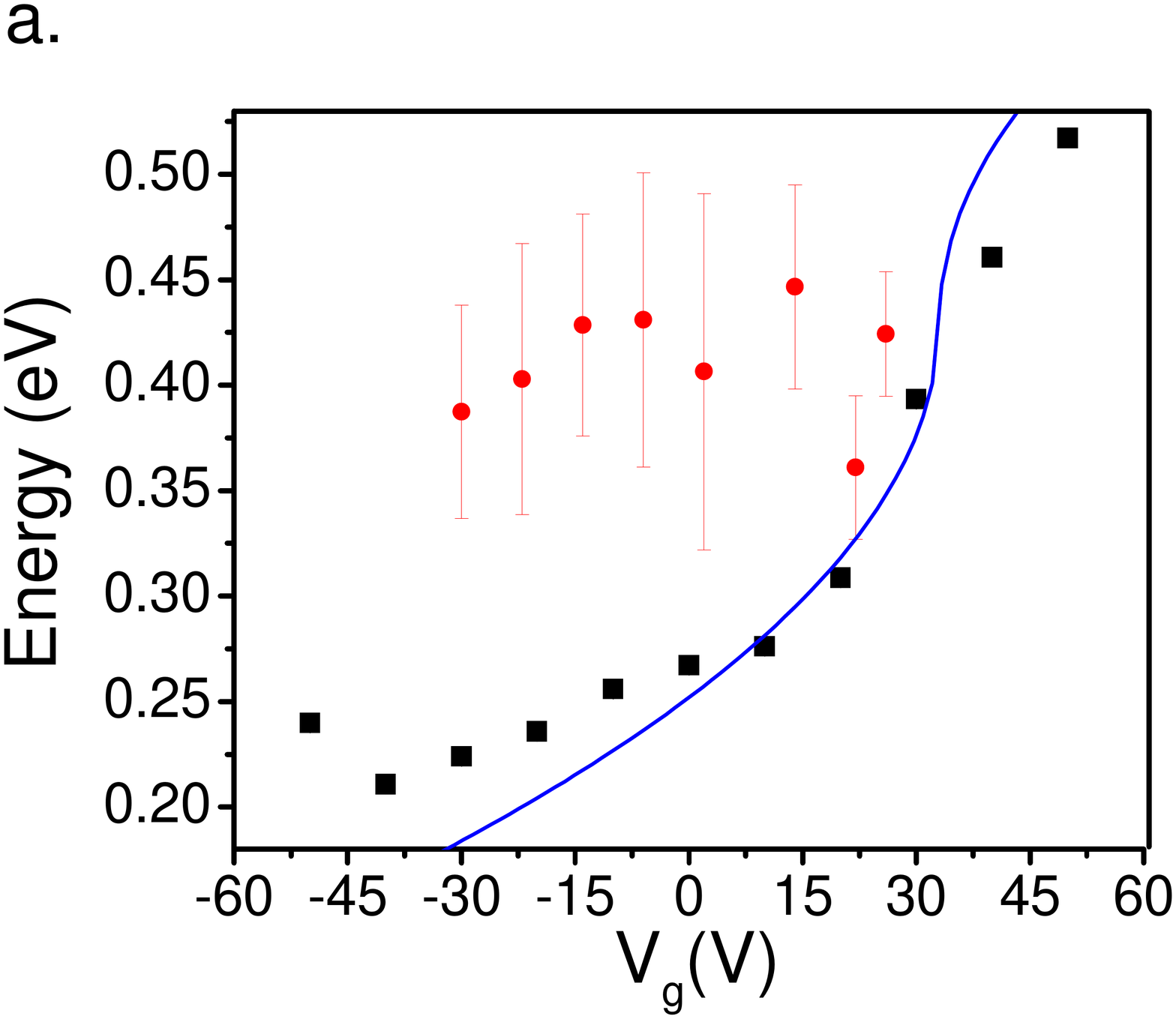}
\includegraphics[width=8cm]{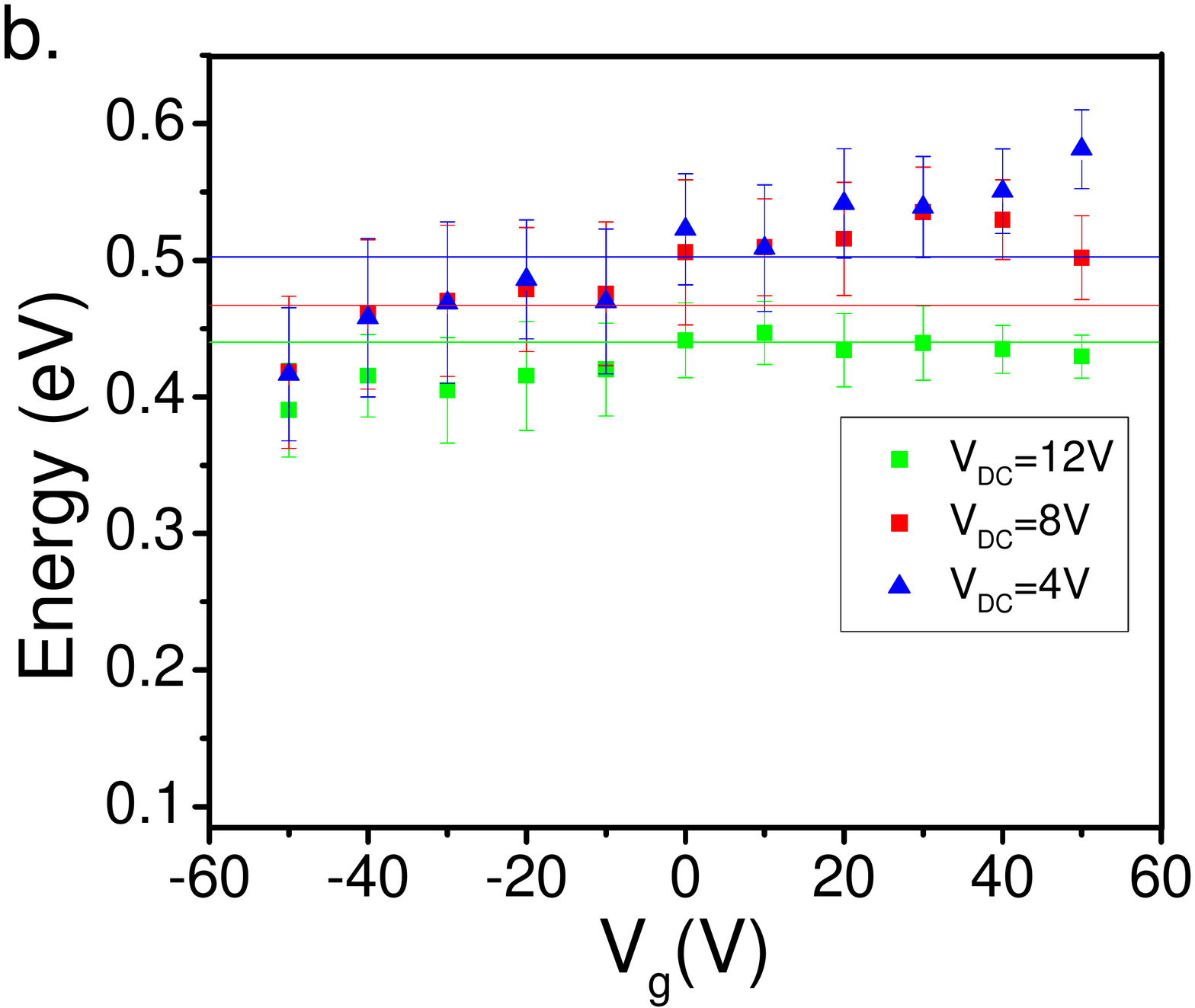}
\end{tabular}
\caption{Energy barrier at different gate voltages for graphene extracted from the fits to the thermionic emission theory in figure \ref{Fig2} in the temperature range 270K-288K (a) (black squares) and in the temperature range 288K-300K (b). In (b) the barrier energy is shown for different bias voltages 4, 8, and 12 V. The continous line in (a) correspond to the modulation of the Fermi energy of graphene with the gate voltage assuming that the graphene neutrality point is at 50V. The continuous lines in (b) correspond to the expected changes in energy barrier due to the Frenkel Poole correction at different bias voltages (see text). The energy barrier at different gate voltages for graphite (extracted at bias voltage 5V and 240$<$T$<$252) is shown for comparison, red circles in figure a).}
\label{Fig3}
\end{center}
\end{figure}

Figure \ref{Fig3} a) shows the modulation of the energy barrier of graphene (black squares) extracted from Fig. 2 in the range 270 K -288 K at low dc voltage (V$_{DC}$=4 V). In a range of 100 V of gate voltage the energy barrier is modulated importantly, from 210 meV to 520 meV. For comparison the solid line indicates the expected change in Fermi energy with gate voltage in monolayer graphene, considering that the barrier height at the charge neutrality point is the energy barrier for graphite (0.43 eV, red circles) and that the charge neutrality point is at V$g$=33V. The fact that the change in barrier height corresponds well with the expected Fermi energy shift in graphene provides good evidence that the device operates as a barristor, and that the pentacene-graphene contact shows little influence of interface states.

Figure 3b shows the dependence of the barrier energy extracted from Fig. 2 in the range 288 K$<T<$ 300 K on gate voltage at three bias voltages. As noted previously, in this temperature range the barrier energy is relatively independent of gate voltage. However we see a clear bias voltage dependence, with lower energies at higher bias voltages.

The lowering of barrier energy with bias voltage suggests the Frenkel-Poole effect in which the mobility of the device is electric field dependent \cite{Sze}, \cite{Berke}
\be\label{FrenkelPoole}
J=qn\mu_0E\exp\Big(-\frac{q(\phi_{FP}-\beta\sqrt{E})}{k_BT}\Big)
\ee
where $\mu_0$ is the zero-field electron mobility, E the electric field (V$_{DC}$/d), d is the thickness of pentacene, $\beta=\big(q/\pi\epsilon\epsilon_0\big)^{1/2}$ and $\epsilon$ the dielectric constant of pentacene. In Fig. 4 we examine the $I(V)$ characteristics at high V $>$ 7 V, plotting ln($I/V$) vs. $V^{1/2}$. The $I(V)$ characteristics show a Frenkel Poole behavior, seen as linear relationship in Fig. 4. Such behavior has been observed in other graphene-pentacene devices \cite{Berke}. We interpret this high temperature, high bias voltage behavior as a $parallel$ current channel limited by the conductance of the pentacene itself, presumably activated due to trapping/de-trapping processes in the pentacene. Eqn. \ref{FrenkelPoole} predicts a bias-voltage-dependent barrier height; we plot the predicted barrier from Eqn. \ref{FrenkelPoole} on Fig. 3b (dashed lines) assuming a zero-bias barrier of $\phi_{FP}=0.588$ eV, which we interpret as the characteristic trapping energy in the bulk pentacene. The observed dependence of the activation barrier on bias voltage, and the independence of the barrier energy on gate voltage, are quantitatively consistent with Frenkel-Poole behavior in the bulk of the pentacene. We note that the Frenkel-Poole channel appears to be in parallel with the conductance channel limited by thermal activation over the graphene/pentacene Schottky barrier. This is reasonable if the Frenkel-Poole conductance channel involves direct tunnelling from graphene into bulk pentacene traps which lie near the graphene Fermi energy, while the Schottky channel involves activation of electrons to the pentacene mobility edge where they conduct readily through the bulk pentacene. Reduction of the parallel bulk Frenkel-Poole conductance by e.g. use of cleaner semiconductor material with fewer charge traps would increase the contribution of the Schottky channel and allow greater gate-voltage modulation of the current.

\begin{figure}[H]
\begin{center}
\includegraphics[width=8cm]{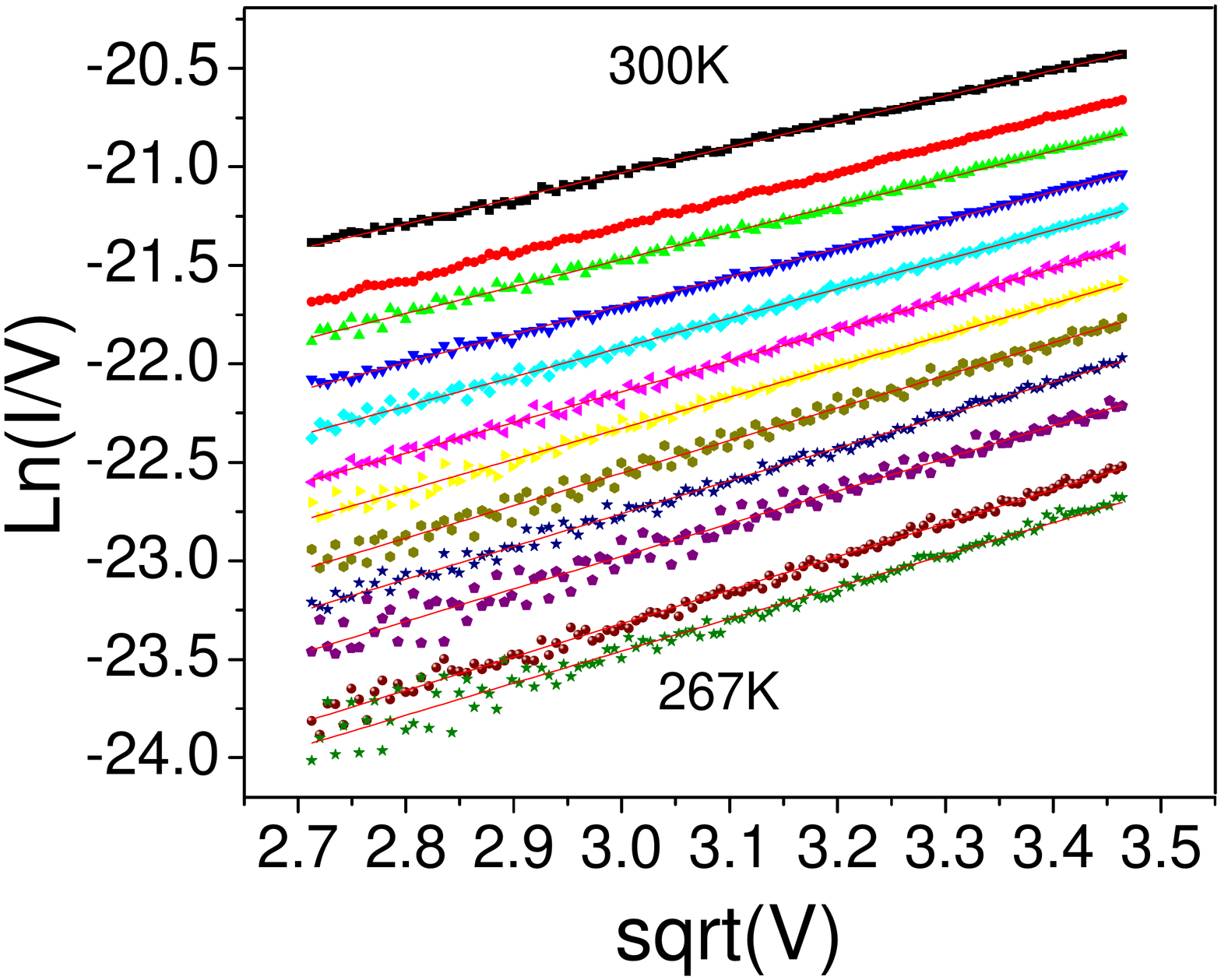}
\caption{Semilog plot of the conductance as a function of the square root of the applied voltage for V$_{DC}>$7V from T=267K to T=300K every 3K for V$_g$=50V.}
\label{Fig4}
\end{center}
\end{figure}

In conclusion, we have shown the feasibility of a thin film barristor made of graphene and pentacene. We observe a gate-voltage controlled modulation of the activation barrier for transport across the graphene/pentacene interface of over 300 meV, which provides strong evidence for a low density of interface states. We expect that our results can be broadly applied to a range of molecular and polymer semiconductors on graphene. The modulation of current through the graphene/organic interface may be of interest in the fabrication of organic transistors, organic light-emitting diodes, and organic solar cells. With modest increases in the range of tunability of graphene's Fermi energy, for example by more efficient electrochemical gates \cite{Yu} or by chemical doping \cite{Liu}, \cite{Ochedowski}, we envision that the interfacial barrier could be reduced to zero, providing highly transparent interfaces for increased efficiency in organic devices.

We acknowledge support of the National Science Foundation under grant DMR-11-05224. M.S.F. acknowledges support from an Australian Research Council Laureate Fellowship. C.O.A. would like to thank Jacob Tosado for providing advise in the pentacene deposition.

\end{document}